\documentclass[hyper]{JHEP3}

\usepackage{amsmath,amssymb,graphicx}
\usepackage{epsfig}
\newif\ifdohyperref
\dohyperreffalse

\dohyperreftrue

\newcommand{\beq}{\begin{eqnarray}}
\newcommand{\eeq}{\end{eqnarray}}

\newcommand{\centeron}[2]{{\setbox0=\hbox{#1}\setbox1=\hbox{#2}\ifdim
                             \wd1>\wd0\kern.5\wd1\kern-.5\wd0\fi \copy0
                             
\kern-.5\wd0\kern-.5\wd1\copy1\ifdim\wd0>\wd1
                             \kern.5\wd0\kern-.5\wd1\fi}}
\newcommand{\ltap}{\>\centeron{\raise.35ex\hbox{$<$}}
                     {\lower.65ex\hbox{$\sim$}}\>}
\newcommand{\gtap}{\>\centeron{\raise.35ex\hbox{$>$}}
                     {\lower.65ex\hbox{$\sim$}}\>}

\newcommand\ZZ{\hbox{\zfont Z\kern-.4emZ}}
\font\zfont = cmss10 
\newcommand{\sfrac}[2]{{\textstyle\frac{#1}{#2}}}

\newcommand{\pd}{\partial}

\ifdohyperref
  \ifx\href\undefined
    \usepackage[hyperindex]{hyperref}
  \fi
  \newcommand\eprint[1]{\href{http://arXiv.org/abs/#1}{[arXiv:#1]}}
  \newcommand\doi[1]{\href{http://dx.doi.org/#1}{[doi:#1]}}
\else
  \newcommand\eprint[1]{[arXiv:#1]}
  \newcommand\doi[1]{[doi:#1]}
\fi

\renewcommand\doi[1]{\unskip}

\title{Warped Domain Wall Fermions}
\author{Tanmoy Bhattacharya$^a$, Csaba Cs\'aki$^b$, Matthew R. Martin$^a$,
Yuri Shirman$^a$, {\rm and} John Terning$^c$ \\
$^a$ Theory Division T-8, Los Alamos National Laboratory, Los Alamos, NM 87545 \\
$^b$ Institute of High Energy Phenomenology, Newman Laboratory, Cornell University, Ithaca, NY 14853 \\
$^c$ Department of Physics, University of California, Davis, CA 95616\\
E-mail: \email{tanmoy@lanl.gov}, \email{csaki@lepp.cornell.edu},  \email{mrmartin@lanl.gov}, 
\email{shirman@lanl.gov}, \email{terning@physics.ucdavis.edu}}

\preprint{\heplat{0503011}}

\abstract{We consider Kaplan's domain wall fermions in the presence of an
Anti-de Sitter (AdS) background in the extra dimension.  Just as in
the flat space case, in a completely vector-like gauge theory defined
after discretizing this extra dimension, the spectrum contains a very
light charged fermion whose chiral components are localized at the
ends of the extra dimensional interval.  The component on the IR
boundary of the AdS space can be given a large mass by coupling it to
a neutral fermion via the Higgs mechanism.
In this theory, gauge invariance can be restored either by taking the
limit of infinite proper length of the extra dimension or by
reducing the AdS curvature radius towards zero.  In the latter case, the
Kaluza-Klein modes stay heavy and the resulting classical theory
approaches a chiral gauge theory, as we verify numerically. Potential
difficulties for this approach could arise from the coupling of the
longitudinal mode of the light gauge boson, which has to be treated
non-perturbatively.}

\begin{document}

\section{Introduction}
\label{sec:introduction}
\setcounter{equation}{0}

Though exact regularizations of chiral gauge theories in the Higgs
phase have long been considered~\cite{early}, the non-perturbative
definition of {\em unbroken} chiral gauge theories has remained a
vexing problem.  The most natural attempts to discretize the theory
on a space-time lattice ran up against the famous Nielsen-Ninomiya no-go
theorem~\cite{NielsenNinomiya} which states that no discretization of
the Euclidean 4D Dirac operator can have the correct free fermion
spectrum and dispersion relation in the continuum limit if it is
translationally invariant, local, and chirally invariant.  Ginsparg
and Wilson~\cite{GW} realized that a possible way out was to violate
the chiral symmetry mildly, {\it i.e.,} the anti-commutator of the
fermion propagator with $\gamma_5$ could be made zero at all non-zero
distances, and explicit realizations of this form have recently been
found~\cite{Neuberger}.  L\"uscher~\cite{Luescher} showed that the
Ginsparg-Wilson relation implied an exact lattice symmetry which
reduced to the chiral symmetry in the na\"\i{}ve continuum limit, but
this symmetry depended on the interactions of the fermion.  Such a
structure allows a consistent chiral projection of the theory, but the
interaction dependence has hindered an explicit discretization of the
projected fermions involved in unbroken chiral gauge interactions.
The current attempts at defining such a theory, therefore, hinge on
directly defining the fermion measure~\cite{Luescher2}, or on fixing
the gauge symmetry on the lattice only to restore it in the
continuum~\cite{Rome,fine}. Obtaining this regularization directly as
a limit of more standard gauge invariant discretizations could
possibly provide better understanding of its nonperturbative aspects,
and is the goal of this paper.

Kaplan~\cite{Kaplan} has shown that fermions in a 5D gauge theory can
have 4D almost-zero modes whose opposite chirality components are
localized at different positions in the 5D bulk.  Since the 4D Dirac
operator defined on the Kaluza-Klein modes of this theory does not
satisfy the assumptions of the Nielsen-Ninomiya no-go theorem, it
would seem that if one could further localize the gauge field around
one of these zero modes, say the left-handed one, then one would have
a chiral gauge theory on the lattice.  Having a gauge interaction
turned on where the left-handed fermion is localized and turning this
interaction off somewhere in between this position and that of the
right-handed fermion, however, has been shown to result in two new
zero modes~\cite{golterman} with opposite chiralities, exactly one of
which couples to the 4D gauge zero mode rendering the theory
vector-like.  It has been noted that one could try to decouple the
extra fermions by adding Yukawa couplings and an appropriate Higgs
vacuum expectation value (VEV), but that taking the VEV large enough to decouple the fermions
would also make the gauge boson heavy~\cite{golterman} and thus result
in a spontaneously broken gauge theory rather than the long sought
after unbroken chiral gauge theory.

Recent developments~\cite{RS,higgsless,higgsless2,otherhiggsless} in the
phenomenology of electroweak symmetry breaking seem to offer a way
out. In theories with an extra dimension, there are gauge boson modes
that vanish at a boundary, and these are not affected by a VEV
localized there.  As a result even in the infinite VEV limit there are
modes of the gauge boson which stay finite in mass, decouple from
the Higgs, and result in unitary scattering without any contribution
from the Higgs~\cite{higgsless,otherunitarity}.  Finally, when the
extra dimension is a warped space, such as 5D Anti-de Sitter (AdS$_5$),
the limit of zero curvature radius makes the lightest of these modes
massless, leaving the rest of the Kaluza-Klein spectrum heavy.  Thus,
it seems we can have our cake: a gauge-breaking fermion mass, and eat
it too: an arbitrarily light gauge boson with no light Kaluza-Klein
modes.  In this paper we will show explicity how to latticize the
extra dimension and discuss how to take a limit of this vector-like
gauge theory that becomes a chiral gauge theory at the classical level.

We start with a brief review of Kaplan's original domain wall fermion
idea in Section~\ref{sec:2}, and review, in the next section, the
arguments that suggest that it is impossible to decouple one of the
localized light modes without either having new light fermions pop up,
or giving a large mass to the gauge boson. Next, we review fermions in
warped extra dimensions, and present the continuum version 
of chiral warped domain wall fermions.  In Section~\ref{sec:5} we
discuss the discretization of the extra dimension and the required
scalings, while in Section~\ref{sec:6} we demonstrate that the
classical discretized theory with fermions is indeed chiral. The main
worry about this proposal is the effect of the longitudinal component of
the gauge field, which becomes strongly coupled near the IR brane.
This is discussed in Section~\ref{sec:7}. Section~\ref{sec:8} contains
comments about a first attempt to latticize the remaining four
dimensions without violating the underlying AdS symmetries. Finally
issues regarding anomalies and instantons are discussed in
Section~\ref{sec:9}.

\section{Review of the Kaplan domain wall fermion proposal}
\label{sec:2}
\setcounter{equation}{0}

Let us start our discussion with an overview of fermions in an extra
dimension, and with it the domain wall approach proposed by
Kaplan.\footnote{See also \cite{otherfermion,ourfermion} for other discussions
of fermions in extra dimensions.} The Lorentz group in 5D is bigger than the 4D Lorentz group,
and the 5D Clifford algebra, by definition, also includes
$\gamma_5$.  An irreducible fermion representation of
the 5D Lorentz group has to contain both chiralities,
and a 5D theory is, therefore, non-chiral, as long as 5D Lorentz
invariance is not broken. A generic 5D fermion action
in flat space can be written as
\begin{equation}
S = \int d^5 x \left(
- i  \bar{\chi}  \bar{\sigma}^\mu \partial_\mu \chi
- i \psi  \sigma^{\mu} \partial_\mu \bar{\psi}
+ \frac{1}{2} ( \psi \overleftrightarrow{\partial_5} \chi
-  \bar{\chi}  \overleftrightarrow{\partial_5} \bar{\psi} )
+ m \left( \psi \chi + \bar{\chi} \bar{\psi} \right)
\right)\,,
\end{equation}
where $\psi$ and $\chi$ are the two-component Weyl spinors
corresponding to the chiral components making up a Dirac spinor. We
will call $\psi$ the left, and $\chi$ the right, chiral component in
this paper.  Since the theory is vectorlike, a 5D ``bulk mass'' is
allowed, and is denoted by $m$ here. The above action simply follows
from writing out the 5D action
\begin{equation}
\int d^5 x \bar{\Psi} (i \Gamma_M \partial^M +m)\Psi\,,
\end{equation}
($M=0,1,2,3,5$) in terms of 4D components
\begin{equation}
\Psi =\left( \begin{array}{c} \chi \\ \bar{\psi} \end{array} \right)\,,
\end{equation}
and using $\Gamma_\mu=\gamma_\mu$, $\Gamma_5=i \gamma_5$
($\mu=0,1,2,3$) in the usual
Dirac basis for the Clifford algebra.

One can distinguish between the left and right chiralities of the
fermions by breaking 5D Lorentz invariance in some way. The mechanism
proposed by Kaplan was to consider a kink generated by a scalar $\phi$
in the extra dimension.  In this kink background there will be a
single zero mode with definite chirality localized around the center
of the kink. To see this, assume that the Lagrangian is of the form

\begin{equation}
\int d^5 x \bar{\Psi} (i \gamma_M \partial^M +\phi(y))\Psi \,,
\end{equation}
where the background $\phi (y)$ is such that $\phi (y)\to -v_0$ for
$y\to -\infty$ and $\phi (y)\to v_0$ for $y\to \infty$, and $\phi
(y_0)=0$. The equation of motion in terms of the two component spinors
will then be:
\begin{eqnarray}
-i \bar{\sigma}^{\mu} \partial_\mu \chi - \partial_5 \bar{\psi} + \phi
(y) \bar{\psi} = 0\,,
\nonumber \\
-i \sigma^{\mu} \partial_\mu \bar{\psi} + \partial_5 \chi + \phi (y)
\chi = 0\,.
\end{eqnarray}

In order to find the 4D modes that solve these equations we write down
the KK decomposition for the 5D spinors:
\begin{eqnarray}
          \label{eq:DiracKK1}
\chi  = \sum_n g_n(y)\, \chi_{n} (x)\,, \\
\bar{\psi} = \sum_n f_n(y)\, \bar{\psi}_n (x)\,,
\label{eq:DiracKK2}
\end{eqnarray}
where $\chi_n$ and $\psi_n$ are two-component 4D spinors which form a
Dirac spinor of mass $m_n$ and satisfy the 4D Dirac equation:
\begin{eqnarray}
\label{eq:Diraceq1}
-i \bar{\sigma}^{\mu} \partial_\mu \chi_{n} + m_n\, \bar{\psi}_n = 0\,, \\
-i \sigma^{\mu} \partial_\mu \bar{\psi}_n + m_n\, \chi_{n} = 0\,.
\label{eq:Diraceq2}
\end{eqnarray}
Plugging this expansion into the bulk equations  we get the following
set of coupled first order differential equations for the wave
functions $f_n$ and $g_n$:
\begin{eqnarray}
          \label{eq:1stOrder1}
g_n' + \phi (y)\, g_n - m_n\, f_n = 0\,, \\
          \label{eq:1stOrder2}
f_n' - \phi (y)\, f_n + m_n\, g_n = 0\,.
\end{eqnarray}
For zero modes $m_n=0$, and we get two decoupled first order equations
which can be immediately solved~\cite{Kaplan,AS}:
\begin{eqnarray}
\label{eq:zeromodes}
& g_0' + \phi (y)\, g_0 =0; \ \ & g_0(y) \propto e^{-\int_{-\infty}^y \phi
(y')dy'} \\
& f_0' - \phi (y)\, f_0  =0; \ \  & f_0(y) \propto e^{\int_{-\infty}^y \phi
(y')dy'}\,.
\end{eqnarray}
If the extra dimension is infinite, then only one of the two zero mode
wave function chiralities will be normalizable, and thus we have
achieved our goal of generating a chiral theory starting from a
totally non-chiral one. In the case of a kink with $\phi \to -v_0$ for
$y\to -\infty$ we find that only the function $g_0$ is normalizable,
so there is a zero mode in $\chi$, while for an anti-kink the
situation would be reversed.

\section{The Golterman-Shamir no-go arguments against a 
chiral domain wall fermion theory}
\label{sec:3}
\setcounter{equation}{0}

In the previous section, we obtained a chiral theory from a completely
vectorlike model.  It, however, does not look like a 4D theory even at
long distances, since there is a continuum of 4D fermion and gauge
boson modes.  In order to make the theory at low energies look like
a 4D theory, we need to consider the theory on a finite interval or
assume boundary conditions (BC's) with similar effects.  Since the
chirality was achieved by pushing one of the modes to infinity, as
soon as we depart from the strictly infinite extra dimension, the
theory will become non-chiral again. One simple example to explain
this is to consider the case when the extra dimension is made finite
by imposing periodic BC's in the extra dimensional coordinate $y$. In
that case $\phi (y)$ is periodic too, so if there is a kink
at $y_0$ there needs to be an anti-kink somewhere else. In this case
the anti-kink will support a zero-mode of opposite chirality (in fact
the modes at the kink and anti-kink will interact and there will not
be any exact zero modes) and the theory will not be chiral.

Let us discuss this issue in more detail in terms of a theory
discretized along the extra dimension but still left in the continuum
limit along the four transverse dimensions. This is commonly
referred to as a theory with a deconstructed extra
dimension~\cite{deconstruction}.\footnote{Other interesting
applications of deconstruction involve attempts to formulate
supersymmetric theories on a lattice.~\cite{susylattice,Erich}} Let us
have as our starting point a theory on a finite interval $0<y<L$,
and with a bulk mass for the fermions $m$. For this case
(\ref{eq:zeromodes}) can still be applied (with $\phi (y)=m$) to find
the two possible zero mode solutions~\cite{KapTait}:
\begin{eqnarray}
&& g_0(y)=e^{-my} \\
&& f_0(y)=e^{my}\,.
\end{eqnarray}
These two possible zero modes of opposite chiralities are localized on
the opposite ends of the extra dimension. Thus we can see that a
simple bulk mass in a finite interval acts exactly like a domain wall
and one does not need to complicate the discussion by involving a
scalar field profile. 

To make the Hamiltonian self-adjoint, one needs to enforce appropriate
boundary conditions.  In the continuum 5D theory one can consistently
impose Dirichlet boundary condition of the form
\begin{equation}
\psi (0)= 0 \qquad \hbox{or} \qquad \psi (L)=0\,,
\end{equation}
(or the same BC for $\chi$). In this case the zero mode for $\psi$
(respectively, $\chi$) would be eliminated, leaving us with a chiral
theory on a finite interval.  A similar condition on the deconstructed
theory would seem to run afoul of the Nielsen-Ninomiya
theorem~\cite{NielsenNinomiya}, and would, thus, be a barrier to
further latticizing the four dimensional slices on the boundaries.  We
therefore do not introduce such boundary conditions and work in the
theory where both zero-modes are present.

An explicit construction for the fermions is given by the following
deconstructed action~\cite{SmithSkiba}:
\begin{equation} 
\sum_{i=1}^N  [ -i \bar{\chi}_i \bar{\sigma}^\mu
\partial_\mu \chi_i -i \psi_i \sigma^\mu \partial_\mu \bar{\psi}_i]
+\sum_{i=1}^{N-1} \frac{1}{a} [\psi_{i+1}(\chi_{i+1}-\chi_i)+
   m a \, \psi_{i+1}\chi_{i+1} ]\,.
\label{eq:Lag1}
\end{equation}
Here $a$ is the lattice spacing, $i$ labels the sites in the
fifth direction, and $Na=L$. The structure of this
action is schematically depicted in Fig.~\ref{fig:flatlattice1}.
We have initially chosen not to add a link between the
first $\psi$ and $\chi$ fields, since for this
set of masses it is quite simple to understand the spectrum of zero
modes.  The construction can then be extended via adding the extra link.  In this
particular choice we have made it is quite clear that there have to be
two exact zero modes in the spectrum. Since $\psi_1$ has no mass term
at all, it is massless. Of the remaining fields there are $N-1$ left
handed and $N$ right handed fields, so one combination of the right
handed $\chi$ fields also needs to be massless. In fact it is easy to
find the zero mode of the mass matrix
\begin{equation}
M= \frac{1}{a} \left( \begin{array}{ccccc} ~~& ~ & \\
-1 & 1+ ma  \\ & -1 & 1+ ma \\ & & \ddots & \ddots \\
&&&-1 & 1+ma \end{array} \right)\,.
\end{equation}
The zero mode is given by
\begin{equation}
\chi_i = (1+ma)^{-i}\,,
\end{equation}
which is just the discretized version of one of the continuum zero mode
wave functions $e^{-my}$.

Of course it is quite unnatural not to add the mass term on the first
site. Using the above analysis of zero modes in the absence of $m
\psi_1 \chi_1$ one can however understand easily the effect of
adding this operator to the Lagrangian in (\ref{eq:Lag1}). In the model
without this term the $\psi$ zero mode is always at the first site
$\psi_1$, while the $\chi$ zero mode is exponentially increasing (for
$m<0$) or exponentially decreasing (for $m>0$) away from
$\chi_1$. Thus adding the $\psi_1 \chi_1$ mass term will totally
remove the zero modes if $m>0$, since in that case it is a mass term
for two zero modes localized almost at the same location. However, for
$m<0$ adding this term will only have a small effect on the $\chi$
zero mode, since that has a very small overlap with $\chi_1$, and so
one still expects an extremely light Dirac particle in the spectrum,
whose mass is exponentially suppressed compared to all the other
modes. This very light Dirac mode (whose $\psi$ component is mostly
$\psi_1$ and whose $\chi$ component is mostly $\chi_N$) will be the
approximate Kaplan domain wall fermion.

\FIGURE[t]{\epsfig{file=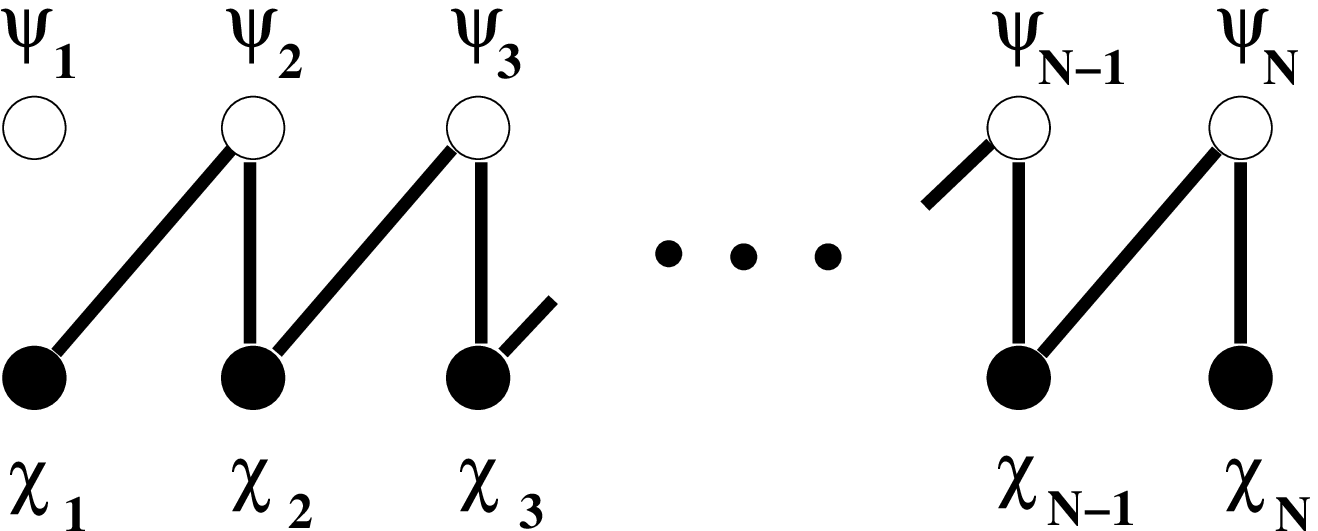,width=.6\textwidth}
\caption{The deconstruction of a fermions in a flat extra dimension.
The empty circles denote left handed two component fermions, while the
filled circles denote right handed ones. A link between two circles
indicates that a mass term is present connecting the two fermions,
either from the bulk mass term, or from the discretization of the 5D
derivative piece.
\label{fig:flatlattice1}}}

Let us now discuss what happens in the presence of a gauge field. We
consider for now a gauge field which does not fluctuate in the fifth direction so
that there is one $SU(n)$ symmetry under which all the fermions transform.
We do not have a chiral gauge theory since there
are two very light fermion modes of opposite chiralities localized at the
two ends of the interval, and both of them couple equally strongly to
the gauge field.

To remedy this situation, two proposals were studied by Golterman
{\it et al.}~\cite{golterman}:

$\bullet$ The gauge field does not propagate everywhere, but only in a
region (called the wave guide) around the first site, this way the second
zero mode does not have a gauge coupling.

$\bullet$ The gauge field is Higgsed at the last site where the
opposite chirality fermion zero mode lives and the second zero
mode gets a mass with some gauge singlet fermion on the last
site.

However, Golterman and Shamir~\cite{GS} (see also~\cite{golterman})
have argued that neither of these possibilities will actually make the
theory really chiral. Their arguments can be summarized as follows.
Let us first consider the case when the gauge field is restricted to a
``wave guide'' that is comprised of the first $k$ sites. This would
mean that the first $k$ fermions need to be thought of as transforming
under a gauge symmetry, while the last $N-k$ would not.  
In order for this to be gauge invariant a charged scalar, $H$, would need to be
associated with the coupling of the charged $\chi_k$ to the uncharged
$\psi_{k+1}$.  So the Lagrangian would be given by
\begin{eqnarray}
&& \sum_{i=1}^k  [ -i \bar{\chi}_i \bar{\sigma}^\mu
D_\mu \chi_i -i \psi_i \sigma^\mu D_\mu \bar{\psi}_i]+\sum_{i=k+1}^N  [
-i \bar{\chi}_i \bar{\sigma}^\mu
\partial_\mu \chi_i -i \psi_i \sigma^\mu \partial_\mu \bar{\psi}_i]+
   \nonumber \\
   &&\sum_{i=1}^{k-1} \psi_{i+1}(\frac{1}{a} \chi_{i+1}- \chi_i)+
\lambda H\psi_{k+1}\chi_k + \sum_{i=k+1}^N
\frac{1}{a} \psi_{i+1}(\chi_{i+1}-\chi_i) +\frac{1}{a} \psi_N \chi_N+
   \sum_{i=1}^N m \psi_{i}\chi_{i} \,.\nonumber \\
\label{eq:Lag2}
\end{eqnarray}

Note, that we have explicitly included a Yukawa coupling constant
$\lambda$ for the term that is controlling the interaction between the
wave guide and the non-gauged part of the lattice.  Golterman and
Shamir have examined the phases of this model for several values of
$\lambda$, and found that the theory is non-chiral in every case. The
simplest possibility is for $\lambda =0$. In this case one can easily
see (see Fig.~\ref{fig:GS}) that the models falls apart into two
disconnected theories. One is the fully gauged wave guide part and the
other is the ungauged part of the domain wall.  Each of these two
parts themselves form a domain wall model exactly as previously, and
each of these will either have zero modes localized at both ends or at
neither end.  Thus the boundary of the wave guide will act as a domain
wall boundary itself. Obviously, nothing different is expected to
happen for small non-zero $\lambda$, as long as the fundamental field
$H$ does not acquire a VEV. For extremely large values of $\lambda$,
on the other hand, the conclusion is very similar to the $\lambda =0$
case.  One can rescale the fields $\chi_k$ and $\psi_{k+1}$ to absorb
this large Yukawa coupling, but in this case their kinetic terms will
tend to zero and the fields will become non-propagating. Thus in the
limit $\lambda \to \infty$ we simply have a theory where the fields
$\chi_k$ and $\psi_{k+1}$ are eliminated, and so we again get two decoupled
domain wall theories like in the $\lambda \to 0$ case, and the theory
will again be non-chiral. Golterman and Shamir have shown that the
general conclusion remains valid for $\lambda ={\cal O} (1)$ as
well. Thus one does not expect a chiral theory unless some of the
fields develop expectation values.

The second possibility that was considered in~\cite{golterman,GS} is
that the scalar field $H$ in (\ref{eq:Lag2}) obtains a VEV, thus
breaking the gauge symmetry at the boundary. This would be welcome
since then the additional zero mode localized at the wave guide
boundary could be eliminated using the opposite chirality fermion
localized on the other side of the wave guide boundary via the term
$\lambda H\psi_{k+1}\chi_k$. The problem with this approach is that
the fermion mass obtained this way will be of the order $m_f \sim
\lambda \langle H\rangle$. However, in this Higgs' mechanism, the
gauge boson will also pick up a mass of order $M_W \sim g \langle H
\rangle$.  To get to an unbroken chiral theory one would like $M_W \ll
m_f$, however their mass ratio is given by $M_W /m_f \sim
g/\lambda$. Since $\lambda$ is an IR free coupling, at low energies
its value will be determined by $g$, and it seems that no hierarchy
between the masses is possible.  Thus it was argued
in~\cite{golterman,GS} that it is not possible to get a chiral gauge
theory from domain wall fermions.

Below we will argue that the situation is different when one is
considering a non-trivial background metric along the extra
dimension. We will show that in this case the scaling of the gauge
boson mass could be different from that of the fermion mass in the
presence of a symmetry breaking VEV on one of the domain wall
boundaries. This will lead to a possibility of recovering a chiral
gauge theory in the limit when the warping (the background curvature
of the extra dimension) is increased to infinity.

\FIGURE[t]{\epsfig{file=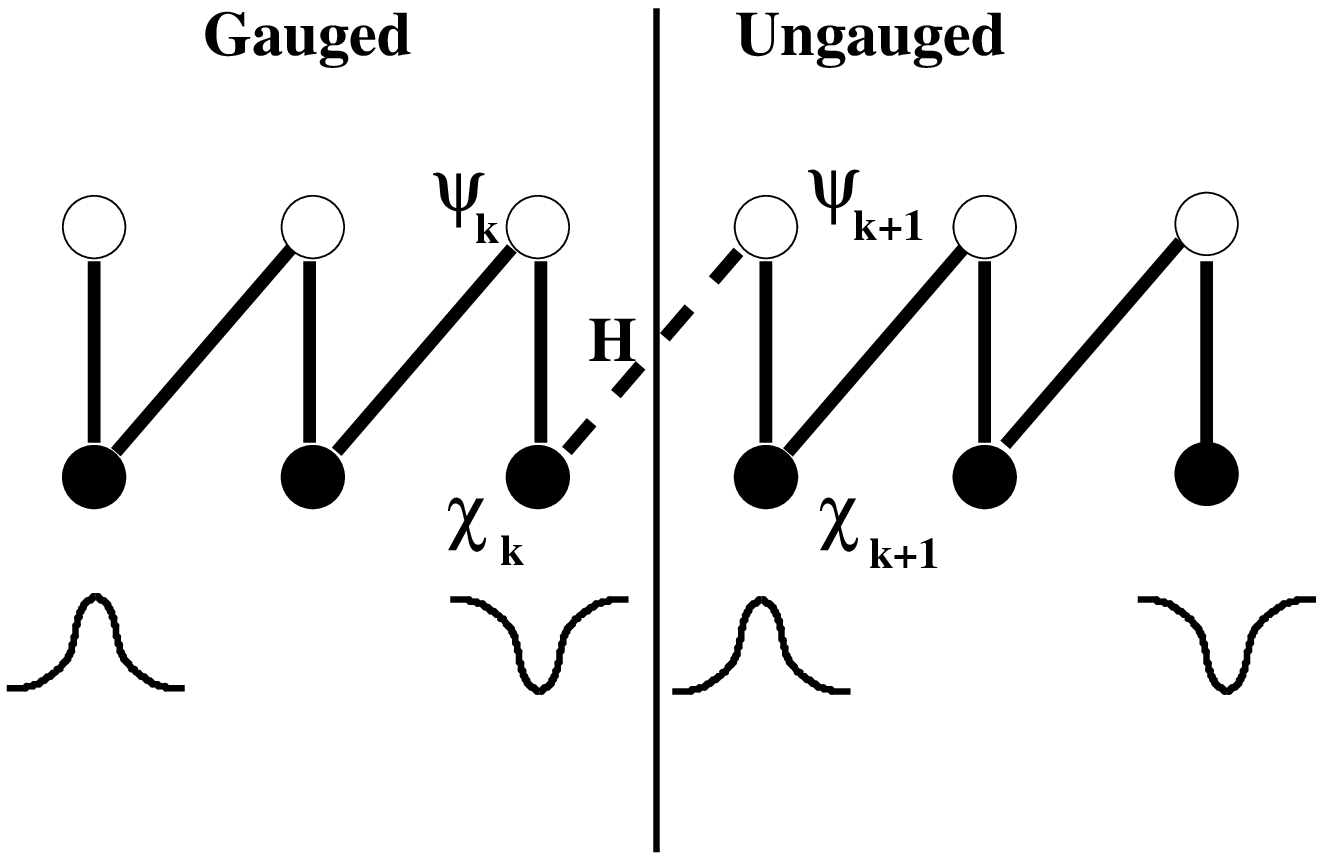,width=.6\textwidth}
\caption{The wave guide model of~\cite{golterman,GS}. The left part of
the model is gauged and called the wave guide, while the right is not gauged. The only
link between the two parts is a Yukawa coupling. Irrespective of the
value of this coupling there will always be an equal number of left
and right handed modes on each part, since the end of the wave guide will act as a
domain wall boundary on its own.
\label{fig:GS}}}

\section{The continuum warped domain wall fermion
theory}
\label{sec:4}
\label{ContinuumWarpedFermions}
\setcounter{equation}{0}

Motivated by the failure of obtaining chiral fermions in a theory with
a flat extra dimension, we will now consider the extra direction to be
curved (``warped''), that is consider a theory in a non-trivial
background metric. For concreteness we will use a five dimensional
anti-de Sitter (AdS$_5$) space given by the background metric
\begin{equation}
ds^2=\left(\frac{R}{z}\right)^2 (\eta_{\mu\nu}dx^\mu dx^\nu -dz^2)\,,
\end{equation}
where $R>0$ is the curvature of the AdS space, and the signature of the
metric is $(+,-,-,-,-)$.  We will consider a finite slice of this AdS
space, that is we restrict $R\leq z\leq R'$. The two boundaries will
be referred to as ``branes'', the one at $z=R$ is usually called the
UV brane, while the one at $z=R'$ the IR brane. The proper distance
between the two branes is given by $R \sqrt{\ln R'/R}$, but, as we
will see, the curvature near the UV brane changes the energy scale that
governs the mass of modes that fluctuate along the fifth dimension to
$1/R'$ instead.  This extra dimensional theory will play the role of the
domain wall theory reviewed in the previous sections.

The background metric has a well known scaling isometry of the form
\begin{eqnarray}
&z &\to \alpha z \nonumber \\
&x_\mu &\to \alpha x_\mu \nonumber \\
&\partial_\mu &\to \frac{1}{\alpha} \partial_\mu\,,
\label{isometry}
\end{eqnarray}
which implies the four dimensional momentum scales are changing along
the extra dimension.  To see this most clearly, consider a four
dimensional scalar theory localized at some slice in the $z$
coordinate:
\begin{equation}
\int d^5x \sqrt{\tilde{g}} \delta(z-z') \left[ \tilde{g}^{\mu\nu}
\partial_{\mu} \phi \partial_{\nu} \phi
- m^2 \phi^2 -\frac{\phi^n}{\Lambda^{n-4}} \right]\,,
\end{equation}
where $\tilde{g}_{\mu\nu}$ is the induced metric on the slice and we
have included one non-renormalizable operator to make the scaling of
the cutoff clear.  After rescaling the field to have a canonical four
dimensional kinetic term, all dimensionful parameters pick up the
appropriate power of the warp factor:
\begin{equation}
\int d^4x \left[ \partial_{\mu}\phi^2 -\left(m \frac{R}{z'}\right)^2
\phi^2
- \frac{\phi^n}{\left(\Lambda \frac{R}{z'} \right)^{n-4}} \right]\,.
\label{eqScalarAction}
\end{equation}
The theory will be invariant under a shift to a different slice and
rescaling the dimensionful parameters, including the cutoff scale of
the theory that will need to become position dependent and decrease as
\begin{equation}
\Lambda (z)= \Lambda (R) \frac{R}{z}\,.\label{eq:cutoffscale}
\end{equation}
This scaling symmetry will be important for preserving the form of the
masses and wave functions of the lightest modes in our theory.  We will
therefore need to reconsider this symmetry carefully when we discuss
the proper lattice theory in Section~\ref{sec:lattice}.

That the dimensionful parameter of the non-renormalizable operator
really is the cutoff in the sense of a regulator for momentum integrals
can be seen by computing one loop perturbative corrections to the mass.
For example, if this scalar theory has a quartic interaction, $\lambda
\phi^4$, then the divergent part of the one loop correction to the mass
is
\begin{equation}
\delta m^2 = \lambda \int^{\Lambda(z)} \frac{d^4p}{(2\pi)^4} \frac{1}{
p^2-\left(m\frac{R}{z'}\right)^2 + i\epsilon}
\sim \lambda \Lambda(z)^2\,.
\end{equation}
The contribution to the mass from this correction only respects the
scaling symmetry if the momentum cutoff has the form determined by the
coefficients in equation~(\ref{eqScalarAction}).

The main reason for considering the extra complication of adding a
background metric is that the presence of this background will significantly modify
the expression for the lowest lying gauge boson mass in the presence
of a scalar VEV.  Unlike the Golterman and Shamir
construction~\cite{golterman}, our gauge field is not held constant
along the extra dimensions.  Let us for example consider a 5D SU(n)
gauge theory and assume that we have added sufficiently many scalars
on the IR brane to completely break the gauge group (for example a
single doublet for SU(2), or two triplets for SU(3)).  The pure
gauge action will be
\begin{equation}
\label{eq:gaugeL}
\int d^5 x\sqrt{g} \frac{-1}{4g_5^2} F_{MN} F^{MN} = \int d^4 x
\int_R^{R'} dz \frac{-1}{4 g_5^2} \frac{R}{z}
[F_{\mu\nu}^2 -2 F_{\mu 5}^2]\,.
\end{equation}
In the limit when the scalar VEV on the IR brane is increased beyond
$1/R'$ the lightest gauge boson mass will approach~\cite{higgsless,Huber:2000fh}
\begin{equation}
M_W^2=\frac{2}{R'^2 \log \frac{R'}{R}} 
\left(1 + {\cal O}\left(\frac{1}{\log\frac{R'}{R}}\right)\right)\,.
\label{eq:Wmass}
\end{equation}
As mentioned above, this mass does not increase beyond a limiting mass given
above as the localized VEV $v$ grows. This is characteristic to any
genuinely extra dimensional theory.  It arises due to the fact that the localized
mass term can push the gauge field away from the brane into the bulk, and
in the limit when the localized mass goes to infinity its effect can
simply be replaced by a Dirichlet boundary condition.  The second 
property, particular to the curved space, is that the mass of the lightest
mode is suppressed by the ``warp factor'' $\sqrt{\log R'/R}$. The more warped
the theory is, the more one is suppressing the lightest gauge boson
mode compared to all the other KK modes whose masses are of order $1/R'$.

Thus, one considers the limit where
\begin{equation}
\frac{1}{R'}\to \infty \,, \qquad \frac{1}{R'^2 \log \frac{R'}{R}} \to 0\,.
\end{equation}
In this limit all the gauge boson KK modes become very heavy, except
for the lightest one which tends to zero.  This is the main
observation of this paper: we will argue that in this limit gauge
invariance will be restored, while one may still be able to use the very large
scalar VEVs on the IR brane to remove from the spectrum any unwanted
light fermions localized there.

Let us summarize next the properties of free fermions in warped space. The
fermion action is given by~\cite{warpedfermions,ourfermion}
\begin{equation}
S = \int d^5 x
\left(\frac{R}{z}\right)^4
   \left(
- i \bar{\chi}  \bar{\sigma}^\mu \partial_\mu \chi
- i \psi  \sigma^{\mu} \partial_\mu \bar{\psi}
+ \sfrac{1}{2} ( \psi \overleftrightarrow{\partial_5} \chi
-  \bar{\chi}  \overleftrightarrow{\pd_5} \bar{\psi} )
+ \frac{c}{z} \left( \psi \chi + \bar{\chi} \bar{\psi} \right)
\right)\,,
\label{eq:warpedaction}
\end{equation}
where $c$ is the bulk mass term in units of the AdS curvature $1/R$.
This can be obtained by evaluating in AdS space the general 5D Dirac
action in curved space
\begin{equation}
          \label{eq:curvedaction}
S = \int d^5 x
\sqrt{g}
\left(
\frac{i}{2} (\bar{\Psi}\, e_a^M \gamma^a D_M \Psi
- D_M \bar{\Psi} \, e_a^M \gamma^a \Psi)
-m \bar{\Psi} \Psi
\right)\,,
\end{equation}
where $e_a^M$ is the vielbein and $D_M$ is a covariant derivative
including the spin-connection.

The bulk equations of motion derived from this action are
\begin{eqnarray}
\label{bulkeq1}
-i \bar{\sigma}^{\mu} \pd_\mu \chi - \pd_5 \bar{\psi} + \frac{c+2}{z}
\bar{\psi} = 0,
   \\
-i \sigma^{\mu} \pd_\mu \bar{\psi} + \pd_5 \chi + \frac{c-2}{z} \chi =
0\,.
\label{bulkeq2}
\end{eqnarray}
The KK decomposition takes its usual form
(\ref{eq:DiracKK1})-(\ref{eq:DiracKK2}), where the 4D spinors $\chi_n
$ and $\bar{\psi}_{n}$ again satisfy the usual 4D Dirac equation with
mass $m_n$ (\ref{eq:Diraceq1})-(\ref{eq:Diraceq2}).  The bulk
equations then become ordinary (coupled) differential equations of
first order for the wavefunctions $f_n$ and $g_n$:
\begin{eqnarray}
          \label{eq:beom1}
& \displaystyle
f^\prime_n + m_n g_n - \frac{c+2}{z} f_n = 0\,,
\\
          \label{eq:beom2}
& \displaystyle
g^\prime_n  - m_n g_n + \frac{c-2}{z} g_n = 0\,.
\end{eqnarray}

For a zero mode, if the boundary conditions allow its presence, these
bulk equations are already decoupled and are thus easy to solve,
leading to:
\begin{eqnarray}
& \displaystyle
f_0  = C_0 \left( \frac{z}{R} \right)^{c+2}\,,
\\
& \displaystyle
g_0 = A_0  \left( \frac{z}{R} \right)^{2-c}\,,
\end{eqnarray}
where $A_0$ and $C_0$ are two normalization constants
of mass dimension $1/2$.

As an example, let us consider the simplest case which is allowed in
the continuum theory, when we make the conventional 
choice~\cite{warpedfermions} of imposing
Dirichlet BC's on both ends:\footnote{For a general analysis of fermion 
boundary conditions see~\cite{ourfermion}.}
\begin{equation}
\psi_{|R^+} = 0
\qquad \hbox{and} \qquad
\psi_{|R^{\prime\, -}} = 0\,,
\label{eq:continuumBC}
\end{equation}
which by~(\ref{bulkeq2}) fixes the BC's for $\chi$.
These BC's allow for a chiral zero mode in the $\chi$ sector while the
profile for $\psi$ has to be vanishing, so we find for an arbitrary
value of the bulk mass $c$ that the zero modes are given 
by~\cite{warpedfermions}:
\begin{equation}
f_0 = 0
\qquad \hbox{and} \qquad
g_0 = A_0 \left(  \frac{z}{R} \right)^{2-c}\,.
\end{equation}
The main impact $c$ has on the zero mode is where it is localized:
close to the UV brane (around $z=R$) or the IR brane (around $z=R'$).
This can be seen by considering the normalization of the fermion wave
functions.  To obtain a canonically normalized 4D kinetic term for the
zero mode, one needs
\begin{equation}
          \label{eq:normM0}
\int_R^{R'} dz
\left(\frac{R}{z} \right)^5 \frac{z}{R} \, A_0^2  \left(  \frac{z}{R}
\right)^{4-2c}
=1
\qquad \hbox{\it i.e.}\qquad
A_0= \frac{\sqrt{1-2c}}{R^c\sqrt{R^{\prime\, 1-2c}-R^{1-2c}}}\,,
\end{equation}
where the first factor in the integral comes from the volume element
$\sqrt{g}$, the $z/R$ factor from the vielbein and the rest is the
square of wave function itself. To conveniently figure out where this
zero mode is localized, we can send either brane to infinity and see
whether the zero mode remains normalizable. For instance, sending the
IR brane to infinity, $R'\to \infty$, the integral (\ref{eq:normM0})
converges only for $c>1/2$, in which case the zero mode is localized
near the UV brane. Conversely, for $c<1/2$, when the UV brane is sent
to infinity, $R \to 0$, the integral (\ref{eq:normM0}) remains
convergent and the zero mode is thus localized near the IR
brane. Repeating this analysis for the other possible zero mode in $f$
we find that this zero mode will be localized on the UV brane if
$c<-1/2$ and on the IR brane for $c>-1/2$.

We can summarize the story of fermion zero modes in a warped metric
as follows: just as in the case of flat space, the localization of
the zero modes depends on the bulk mass parameter $c$. The main
difference is that the presence of the background curvature
effectively acts as a mass term itself, and where the right handed zero
mode, $\chi_0$, is localized depends on the sign of $c-1/2$, while the
localization properties of the left handed mode, $\psi_0$, depend on
the sign of $c+1/2$. Picking appropriate values of $c$ one can arrange
for the different chiralities to be localized on the different branes, just as
in the flat space case. For example for $c=-1$ the
left handed mode is localized on the UV bane and the right handed on
the IR brane.

We have now every ingredient needed to construct the continuum version
of the warped domain wall fermion theory.  We will consider an $SU(n)$
gauge theory in AdS space as above, with fermions in the bulk.  Because
of the Nielsen-Ninomiya theorem, we will not be able to impose the
boundary conditions~(\ref{eq:continuumBC}).  However, we still
choose the bulk mass parameter $c$ such that the two zero modes of
opposite chiralities are localized on the different branes. We then
add several Higgs scalars $H_i$ on the IR brane to break the gauge
invariance.  As discussed above this will result in a gauge boson mass
(\ref{eq:Wmass}) which can still be made small by adjusting the
curvature scale of the bulk. At the same time we can add some
left handed neutral fermions $S_i$ and right handed neutral fermions
$\bar{S}_i$ ($i=1,\ldots ,n$) on the IR brane to the theory. We can
use the scalar to add a Yukawa coupling between the singlet fermions
on the IR brane and the bulk fermions:
\begin{equation}
{\cal L}_{IR}= \sum_i \left( H_i \psi (R') S_i + H_i^*\chi (R') \bar{S}_i +
M_{S_i} (S_i S_i +\bar{S}_i \bar{S}_i )+\ \hbox{h.c.} \right)
\end{equation}
Note, that we have added some Majorana mass terms for the {\em
singlet} fermions on the IR brane. These are necessary to get an odd
number of zero modes in the theory. The effect of these additional mass terms on
the IR brane will be to give a large mass, of order $1/R'$, to the zero
mode localized close to the IR brane, but not affect the zero mode
localized at the UV brane. Thus at the classical level the spectrum of
the theory is expected to be chiral.

\section{Discretization of the 5th direction}
\label{sec:5}
\setcounter{equation}{0}

To study the theory discussed in the previous section, we will
deconstruct~\cite{deconstruction} 
the fifth dimension of this gauge theory in AdS$_5$ --- this
will give us a description with $N$ 4D slices.  We begin
with the classical Lagrangian (\ref{eq:gaugeL}).  It is convenient to
choose a lattice spacing along the $z$ direction which preserves the
scaling symmetry (\ref{isometry}) of the continuum theory:
\beq \delta
z_i = a z_i ~~~ z_i=(1+a)^{i-1}z_1\,,
\label{eq:discr1}
\eeq
where $a\ll 1$ is a dimensionless number. Since the 5D
theory may not be renormalizable, we do not envisage taking the limit
$a\to0$.  We will, nevertheless, keep it small to stay close to the
continuum classical theory and make qualitative use of various 5D
continuum results.  (We will see in section~\ref{sec:6}
that qualitative changes occur in the behavior
of the fermion wave functions as $a$ approaches one.)  It is also convenient 
to define a ``local warp factor'' between two neighboring 4D
slices given by  
\beq
w=\frac{z_{i}}{z_{i+1}}=\frac{1}{1+a}\,.
\label{eq:discr2}
\eeq
This allows us to write a convenient relation between the locations of the
branes in the continuum description and the lattice parameters $w$ and
$N$
\beq
R^\prime = Rw^{-N+1}\,.
\label{eq:discr3}
\eeq

The deconstructed Lagrangian for the 4D gauge fields then takes the form
\beq 
\sum_{i=1}^N \frac{-aR}{4g_5^2} (F_{\mu\nu}(i))^2 + \sum_{i=1}^{N-1}
\frac{aR}{2g_5^2} \frac{(A_\mu(i+1)-A_\mu(i))^2}{a^2z_i^2} + \ldots\,,
\eeq
where, for brevity, we suppress the 4D position
from the arguments.  The second sum gives mass terms for the
4D gauge fields and arises from discretizing $(\partial_5
A_\mu)^2$ in $F^2_{\mu 5}$. Obviously this Lagrangian describes a
product gauge theory (a 4D gauge group is clearly associated with
each 4D slice) with the gauge couplings defined by
\beq 
\frac{1}{g_i^2}=\frac{aR}{g_5^2}\,.  
\eeq 
The mass terms for the
gauge fields break the product gauge group to a diagonal subgroup with
the gauge coupling given by 
\beq
\frac{1}{g_4^2}=\frac{N}{g_i^2}=\frac{NaR}{g_5^2}\,. 
\eeq
Once the
Higgs boson charged under the N'th gauge group obtains a large VEV, $v
=\frac{1}{aR} w^{N-1}$, the gauge boson mass matrix
becomes~\footnote{The mass spectrum and eigenvectors of this mass
matrix will remain essentially unchanged if the Higgs VEV is much
larger.}
\begin{equation}
\label{eq:gaugemass}
   \frac{1}{a^2 R^2 }
\left[ \begin{array}{cccccc} 1&-1\\ -1&1+w^2&-w^2\\ &-w^2&w^2+w^4&-w^4\\
&&&\ddots \\ &&&-w^{2N-6} &w^{2N-6}+w^{2N-4}& -w^{2N-4}\\
&&&&-w^{2N-4}&w^{2N-4}+w^{2N-2} \end{array} \right]\,.
\end{equation}
Notice that in the discretized description the radius of curvature
appears as an explicit mass parameter  (in the combination $aR$) in
the prefactor. 

Eigenvalues and eigenvectors of (\ref{eq:gaugemass}) are similar to
masses and wavefunctions of the KK modes in the continuum description
if the dimensionless lattice spacing $a$ is sufficiently small.  In
order to reproduce the mass of any given mode we need to be able to
sample the oscillations of the corresponding eigenvector.  Despite the
curvature, there are still $j$ oscillations for the KK mode labeled by
$j$, which have a period of approximately $R'/j$ in the $z$
coordinate.  The largest lattice spacing is at the IR brane and should
be less than this period in order to reproduce the continuum
expressions.  Therefore, the lightest ${\cal O}(1/a)$ modes are
expected to approximate the corresponding continuum modes.  In our
numerical tests, we will choose a fixed lattice spacing: $a =
\frac{1}{10}$. 

Thus, in terms of the lattice parameters, we immediately see that, to
the leading order in $a$, the KK mass scale is given by
\beq
\label{eq:mKK}
m_{KK}^2 \approx \frac{1}{R'^2} \simeq \frac{w^{2N}}{R^2}
\eeq
while the mass of the lightest gauge boson is
\begin{equation}
\label{eq:mW}
M_W^2 \approx \frac{2}{R'^2 \log R'/R}
\left( 1 + {\cal O}\left( \frac{1}{\log\frac{R'}{R}}\right) \right) \simeq
\frac{-2 w^{2N}}{R^2 N \log w}.
\end{equation}

Our deconstruction has been performed so far in the classical
theory. Because $1/Ra$ sets the mass scale at the UV brane, and
(\ref{eq:cutoffscale}) requires us to scale our cutoff in a position
dependent way, we should reinterpret the classical theory as the
theory with a varying 4D cutoff $v_i = v_1 w^{i-1}$, with $v_1=1/(Ra)$.  To
preserve the scale invariance of the continuum, we define
\beq
\frac{1}{g_i^2(v_i)}=const.\, =\frac{1}{g_1^2 (v_1)}\,.\label{eq:thdef}
\eeq
The structure of the mass matrix (\ref{eq:gaugemass}) used in the
classical description, then, remains unchanged under renormalization.

With the above definition of the theory, the coupling of the diagonal
group (at the fixed infrared scale $\mu$) is given by~\cite{RSdeconstruct,KS}
\beq
\label{eq:RG}
\frac{1}{g_4^2(\mu)}=\frac{N}{g_1^2(v_1)} + \frac{b}{8\pi^2} \ln 
\frac{\mu}{v_1}+\ldots
=\frac{NaR}{g_5^2}+\frac{b}{8\pi^2}\ln(aR\mu)+\ldots\,.
\eeq

The above formula can be easily derived in the limit $v_i\gg v_{i+1}$
where the separation of scales allows one to integrate out heavy
fields one at a time matching the coupling of a product theory with
$N+1-i$ gauge groups above the scale $v_i$ to that of that of the
effective description with $N-i$ gauge groups below $v_i$. 
One loop evolution of the gauge coupling in the general case of
arbitrary $w=v_{i+1}/v_i$ 
is given in \cite{KS}, but its leading $N$ behavior is captured by
(\ref{eq:RG}). 
From this formula for the running of the coupling we 
can find the one loop expression for $\Lambda_{QCD}$
\begin{equation}
\Lambda_{QCD}=v_1 e^{-\frac{8\pi^2N}{b g_1^2}},
\label{eq:lambda}
\end{equation}
where $g_1$ is understood to be evaluated at the scale $v_1$. 
Our goal is to remove the four dimensional cutoffs $v_i$ in such a way
that the KK tower decouples while the low energy physics is kept fixed 
and the lightest gauge field becomes massless. That is, in the 
$N\rightarrow \infty$ limit, we require
\beq
\frac{m_{KK}}{\Lambda_{QCD}}= a e^{-N(a-\frac{8\pi^2}{bg_1^2})} 
\rightarrow \infty\,,~~~~
\frac{m_{W}}{\Lambda_{QCD}}=\sqrt{\frac{a}{N}} e^{-N(a-\frac{8\pi^2}{bg_1^2})} 
\rightarrow 0\,,
\eeq
where the ratios were obtained using (\ref{eq:mW}), (\ref{eq:mKK}) and 
(\ref{eq:lambda}). One way of achieving this limit is by holding $a$
and the combination 
\begin{equation}
{\cal K} \equiv \frac{a}{N^\frac{1}{4}}\,
e^{-N\left(a - \frac{8\pi^2}{bg_1^2}\right)}
\end{equation}
constant by adjusting $g_1$ as we take $N$ large. Then we may write
the limits we want as: 
\begin{eqnarray}
\frac{m_{KK}}{\Lambda_{QCD}} &=& N^{+\frac{1}{4}}\, {\cal K} \to \infty 
\\
\frac{M_W}{\Lambda_{QCD}} &=& N^{-\frac{1}{4}} \, \frac{{\cal K}}{ 
\sqrt{a}} \to 0\,.
\end{eqnarray}
With this scaling, we find that at large $N$
\begin{equation}
\frac{b g_1^2}{8 \pi^2} \to \frac{1}{a}.
\label{GaugeCouplingScaling}
\end{equation}
Alternatively, we can find a relation for the 5D coupling at the
scale $v_1$:
\begin{equation}
\frac{g_5^2(v_1)}{R} \to \frac{8 \pi^2}{b}\,.
\label{5dGaugeScaling}
\end{equation}
Since the large $N$ limit is taken while holding $a$ fixed, it does not
correspond to the continuum 5D theory. On the other hand to ensure
that the deconstructed description gives a good approximation to the
continuum, $a$ needs to be small, as explained above. 
This implies that one can achieve the desired scaling  
only if the individual gauge group expansion parameters
at the local cutoffs, $v_i$, are large as given 
in (\ref{GaugeCouplingScaling}).\footnote{%
If we attempted to take $a\rightarrow 0$, we would find
$g_i^2(v_i)\rightarrow \infty$, reflecting strong coupling in the 5D
continuum description. On the other hand, keeping $g_i^2(v_i)$ small
requires $a\sim \mathcal{O}(1)$ and even low lying KK modes are not
reproduced in the deconstructed theory.}

\section{The deconstructed warped domain wall fermion theory}
\label{sec:6}
\setcounter{equation}{0}

Next we will discuss the deconstruction of the fermion Lagrangian 
(\ref{eq:warpedaction}). First
we rewrite the continuum action in the form
\begin{equation}
S = \int d^5 x
\left(\frac{R}{z}\right)^4
   \left(
- i \bar{\chi}  \bar{\sigma}^\mu \partial_\mu \chi
- i \psi  \sigma^{\mu} \partial_\mu \bar{\psi}
+ ( \psi \partial_5 \chi +
\partial_5 \bar{\chi}  \bar{\psi} )
+ \frac{c-2}{z} \left( \psi \chi + \bar{\chi} \bar{\psi} \right)
\right)\,,
\end{equation}
which is equivalent up to boundary terms to the action in 
(\ref{eq:warpedaction}). Using the discretization
outlined in (\ref{eq:discr1})-(\ref{eq:discr3}) we 
can write the deconstructed form of this action as
\begin{eqnarray}
S&=& \sum_{i=1}^N aR w^{3i} \left[ -i \bar{\chi}_i
\bar{\sigma}^\mu
\partial_\mu \chi_i -i \psi_i \sigma^\mu \partial_\mu 
\bar{\psi}_i\right] \nonumber \\ &&
+\sum_{i=1}^{N-1} aR w^{3i+3} \left[\frac{w^{i+1}}{a R} 
\psi_{i+1}(\chi_{i+1}-\chi_i)+
\frac{w^{i+1}(c-2)}{R} \psi_{i+1}\chi_{i+1} + \textrm{h.c.} \right]\,.
\label{eq:massmatrix}
\end{eqnarray}
The leading factor of $w^{3i}$ comes from four factors 
of $\frac{R}{z}$ in the determinant of the metric and 
one factor of $z$ in the discretization of the measure, $dz$.
Note, that we have again chosen a discretization where we do not add any 
mass terms for the $\psi_1$. This is so that we can easily identify
the zero modes in the theory.  Later on we will add the
appropriate mass term for this field as well. To get canonically
normalized fields we reabsorb factors of $\sqrt{aR}w^{3i/2}$
into the fields $\chi_i, \psi_i$. The action is then given by
\begin{equation} \sum_{i=1}^N  \left[ -i \bar{\chi}_i \bar{\sigma}^\mu
\partial_\mu \chi_i -i \psi_i \sigma^\mu \partial_\mu 
\bar{\psi}_i\right]+
\sum_{i=2}^N \frac{w^i}{aR} \left[ (1+(c-2)a) \psi_i \chi_i 
-w^{3/2}\psi_i \chi_{i-1} + \hbox{h.c.}
\right]\,.
\end{equation}

The mass matrix then looks like
\begin{equation}
M=\frac{1}{a R} \left[ \begin{array}{ccccc} 0 \\ -w^{7/2} & \alpha w^2 \\ & 
-w^{9/2} & \alpha w^3
\\ & & \ddots & \ddots \\ & & & -w^{N+3/2} & \alpha w^N \\
\end{array}\right]
\label{eq:SmallFermionMassMatrix}
\end{equation}
where $\alpha \equiv (1+(c-2) a)$. We can see from this mass matrix that 
there is a trivial zero mode given by $\psi_1$, and since the number
of left and right handed  fermions are equal there also has to
be another zero mode among the $\chi$ fields. This additional zero mode 
can be found by looking at the above mass matrix:
\begin{equation}
w^{3/2} \chi_i= \alpha \chi_{i+1}\,.
\end{equation}
This wave function for the zero mode is, then,
\begin{equation}
\chi_i = \left( \frac{w^{3/2}}{\alpha}\right)^i 
\sim \left[\left(1-(c-2)a\right)\left(1-\frac{3}{2}a\right)\right]^i 
\sim \left(1-\left(c-\frac{1}{2}\right)a\right)^i
\sim \left(w^{c-1/2}\right)^i\,.
\end{equation}
This clearly is a discretized form of the continuum function
$z^{1/2-c}$, except when $a>1/(2-c)$ and the wave function begins
to oscillate.  The factor of $z^{3/2}$ between this and the continuum
zero mode wave function $z^{2\pm c}$ comes from the rescaling we
performed to get canonical kinetic terms in this section.  Again,
where this wave function is localized depends on the value of $c$.

In order for the zero mode to be localized on the IR brane we need to
pick $c<1/2$. In this case the two zero modes will be spatially well
separated, and adding a mass term $\psi_1 \chi_1$ into the action will
not significantly modify the spectrum. It will also have the effect of
slightly broadening the $\psi$ zero mode, which now will have a wave
function that is approximately the discretization of $z^{1/2+c}$. We
will eventually pick $c<-1/2$ in order for the two zero modes be
spatially separated even after adding the missing term into the mass
matrix in (\ref{eq:massmatrix}).

In order to get to the final mass matrix of the deconstructed warped
domain wall fermion theory, we need to take into account the extra
gauge singlet fermions added on the IR brane that can provide a mass
in the presence of a Higgs VEV. As stated before, if there are only
Dirac masses in the theory a chiral spectrum will never
emerge. However, for the extra localized {\em singlets} one can add a
Majorana mass (of the same order as the masses involving $\psi_N,
\chi_N$), which we will see is sufficient to make the spectrum of
the theory chiral.  Thus we extend the Lagrangian to
\begin{eqnarray} && \sum_{i=1}^N  \left[ -i \bar{\chi}_i 
\bar{\sigma}^\mu
\partial_\mu \chi_i -i \psi_i \sigma^\mu \partial_\mu \bar{\psi}_i
+ \frac{w^i}{aR} (1+(c-2)a) \psi_i \chi_i \right]-
\sum_{i=2}^N \frac{w^{i+3/2}}{aR}\psi_i \chi_{i-1} + {}\nonumber \\ &&
\qquad  m_S \psi_N S +m_{\bar{S}} \chi_N
\bar{S} +m_M (S^2 + \bar{S}^2) +m_D S\bar{S} + \hbox{h.c.}\,,
\end{eqnarray}
where the extra mass terms $m_S,m_{\bar{S}},m_M$ and $m_D$ are assumed
to be of order $\frac{w^N}{aR}$, which is appropriate for a mass term
on the IR brane (last site).  Since this is no longer a pure Dirac
structure, the mass matrix now has to be written in a $(2N+2)\times
(2N+2)$ form:
\begin{equation}
M_{full}= \left[ \begin{array}{cccc|cc|cccc}
& & & & & & & & \\
& & & & & & & & \tilde{M} & \\
& & & & & m_S & \\ \hline
& & & & m_M & m_D & m_{\bar{S}} \\
& & & m_S & m_D & m_M & & & &  \\ \hline
& & & & m_{\bar{S}} & & & & & \\
& & \tilde{M}^\dagger & & & & & & & \\
& & & & & & & &  \end{array} \right]
\label{eq:fullfermion}
\end{equation}
The mass matrix $\tilde{M}$ in some of the off-diagonal blocks has
the form of the mass matrix, $M$, from
(\ref{eq:SmallFermionMassMatrix}), but with the addition 
of a mass term, $\alpha w$, linking $\psi_1$ and $\chi_1$.
We have numerially verified that the spectrum arising from this mass
matrix is indeed in accordance with our expectations. In
Fig.~\ref{fig:spectrum} we show the dependence of the first two
fermion and gauge boson eigenvalues as a function of the warp factor.
\FIGURE[t]{\epsfig{file=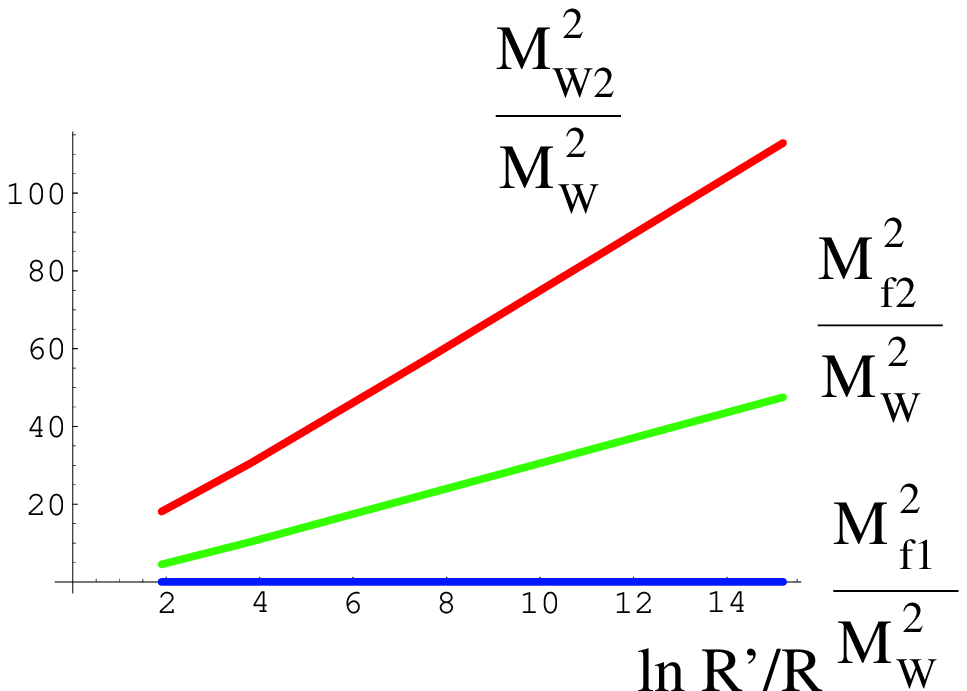,width=.6\textwidth}
\caption{The mass spectrum of fermions and gauge bosons from
\ref{eq:fullfermion} and \ref{eq:gaugemass} as a function of the warp
factor. Here we have chosen $c=-1$ and $a=\frac{1}{10}$, so that
$\ln\frac{R'}{R}$ grows with $N$.  For this plot we used values of $N$
between $20$ and $160$.  The bottom curve shows the ratio of the mass
squares of the lightest fermion vs. the lightest gauge bosons,
confirming that the lightest fermion is essentially massless. The
middle curve is the ratio of the mass squares of the next-to-lightest
fermion to the lightest gauge boson, confirming that the mass of this
fermion increases with the warp factor, and finally the top curve
shows the ratio of the mass squares of the next-to-lightest
vs. lightest gauge bosons.
\label{fig:spectrum}}}

\section{A Strongly Coupled Scalar Mode}
\label{sec:7}
\setcounter{equation}{0}

We have seen that by treating the gauge field dynamically in the fifth
direction and working in a strongly warped AdS space, the gauge symmetry may be
restored in the appropriate limit.  In addition, we have shown that the
light fermion localized near the IR brane may be given a KK scale mass
and decoupled from the theory.  This classical construction is then a
chiral gauge theory on a lattice, at least in the deconstruction
picture we have used so far.  In the absence of any large coupling
constants we would not expect quantum corrections to bring about
qualitative changes to this construction.

However, our construction is based on having an independent gauge
group at each slice along the fifth dimension. Furthermore, as has
been shown in Section~\ref{sec:5}, individual groups are strongly
coupled in the large $N$ limit. Since the Yukawa couplings of the
light fermions to a Goldstone mode of the lightest gauge field are
determined by the gauge couplings and the wave-function overlaps,
these Yukawa's may become large.

Indeed, we are forced to introduce a new degree of freedom on every
four dimensional lattice which is the radially frozen Higgs or, in the
five dimensional language, the fifth component of the gauge field: $A^5$.
As we show below the angular part of this radially frozen Higgs, which
is a would-be Goldstone boson, becomes strongly coupled to the
fermions near the IR brane.  We might worry that these large Yukawa
couplings play the same role as the infinite Yukawa coupling
in~\cite{GS} and push the otherwise massive fermion away from the
brane thereby decoupling it from the brane-localized singlets and
making it light again.  In this section, we calculate these Yukawa
couplings and consider some of the possible implications for the wave
function of the lightest mode and its mass. We will see that in
contrast to~\cite{GS} (where the large, sudden change in Yukawa
coupling was responsible for the modification of the fermion
wave-function) the dependence of Yukawa's on the extra dimensional
coordinate in our case will be smooth.  Whether this difference
suffices to keep all the right handed charged fermions massive will
require a lattice calculation, but we proceed with perturbative
estimates, which are necessarily qualitative, below.

\subsection{Calculation of the Yukawa}
We will again make use of the continuum results from Higgsless models
in $AdS$.  For each KK mode of the gauge field, there is a
corresponding Goldstone mode which is eaten to become the massive
longitudinal component of the gauge KK mode.  Only the Goldstone mode
corresponding to the lightest gauge mode is becoming light as the
symmetry is restored, so we don't expect significant effects from the
other modes.  The coupling of, and mass term for, the gauge field are:
\begin{equation}
g_5 \int_R^{R'} dz \left( \frac{R}{z}\right)^4 A_\mu (z) [
\psi \sigma^\mu \bar{\psi}+\bar{\chi} \bar{\sigma}^\mu \chi ]
+\frac{1}{2} \int_R^{R'} dz \frac{R}{z} M_W^2 A_\mu^2\,.
\end{equation}
As the mass $M_W\to0$, the longitudinal part of the 4D gauge boson
behaves like a scalar Goldstone mode, $\phi$, from which it arose:
\begin{equation}
A_{L\mu} (z) \to \frac{f(z)}{M_W} \partial_\mu \phi\,,
\end{equation}
where $f(z)$ is the wave function in the extra dimension of the
lightest mode of the gauge boson and the normalization is chosen to
make the kinetic term of the Goldstone mode canonical.  By using the
above expressions in the action and integrating by parts with respect to
$\partial_\mu$, we have:
\begin{equation}
g_5 \int_R^{R'} dz \left( \frac{R}{z}\right)^4 \frac{f(z) \phi}{M_W} 
\partial_{\mu}
[\psi \sigma^\mu \bar{\psi}+\bar{\chi} \bar{\sigma}^\mu \chi ]
+\frac{1}{2} \int_R^{R'} dz \frac{R}{z} f^2 \left(\partial_\mu\phi\right)^2\,.
\end{equation}
We can make use of the equations of motion for the fermions,
equations~(\ref{bulkeq1}) and~(\ref{bulkeq2}), and integrate by
parts again, now with respect to $\partial_z$, to get the following
form for the action:
\begin{equation}
ig_5 \int_R^{R'} dz \left( \frac{R}{z}\right)^4 \partial_z f(z)
\frac{\phi}{M_W} (\bar{\chi}\bar{\psi}-
\psi \chi)
+ \frac{1}{2} \left( \int_R^{R'} dz \frac{R}{z} f^2\right) \left(
\partial_\mu \phi \right)^2\,.
\end{equation}
Let us emphasize that the Goldstone mode is a four dimensional degree
of freedom with canonical normalization.  The expressions for the mass and
wave function, to leading order in $\frac{1}{\ln(R'/R)}$, are:
\begin{equation}
M_W^2 =\frac{2}{R'^2 \log \frac{R'}{R}}, \ \ f(z)= \frac{1}{\sqrt{R
\log \frac{R'}{R}}}\left(1-\frac{z^2\log\frac{z}{R}}{R'^2 \log
\frac{R'}{R}}\right)\,.
\end{equation}
Putting all of it together we get the coupling of the fermions to the
Goldstone mode:
\begin{equation}
i \sqrt{2} \frac{g_5}{\sqrt{R}} \int_R^{R'} dz \left( 
\frac{R}{z}\right)^4 \phi
\frac{z}{R'}
\frac{\log{\frac{z}{R}}}{\log{\frac{R'}{R}}}
(\bar{\chi}\bar{\psi}-
\psi \chi)\,.
\end{equation}
If we take into account the fact that the same factor of $\left(
\frac{R}{z}\right)^4$ appears in the kinetic term, the effective
$z$-dependent Yukawa coupling for canonically normalized fields looks
like 
\begin{equation}
y(z)= \sqrt{2} \, \frac{g_5}{\sqrt{R}} \,
\frac{z}{R'} \, \frac{\log{\frac{z}{R}}}{\log{\frac{R'}{R}}}\,.
\end{equation}
In light of the discussion leading to (\ref{eq:thdef}), the bare
five dimensional gauge coupling is to be evaluated at the cutoff
scale, $\frac{1}{Ra}$, but we wish to hold the low energy coupling
fixed.  From the scaling arguments leading to
equation~(\ref{5dGaugeScaling}), the Yukawa coupling becomes
\begin{equation}
y(z)= \frac{4 \pi}{\sqrt{b}} \,
\frac{z}{R'} \, \frac{\log{\frac{z}{R}}}{\log{\frac{R'}{R}}}\,.
\label{eq:yukawa}
\end{equation}

This number, though finite, may be large, so we need to
consider how it affects the wave function of the fermion mode which we
are trying to remove by using the mass term on the IR brane.  It is
important to note that (\ref{eq:yukawa}) is substantially different
from the result in~\cite{GS}. There the Yukawa coupling was localized
at a single site, and it was possible to do either a perturbative or a
strong coupling expansion. Here the Yukawa is smoothly changing in
space between zero and a fixed value of order unity so that, in
this setup, it is not possible to apply either the weak or strong
coupling expansions.  A lattice simulation is necessary to really
decide whether or not there will be additional light fermions.

\subsection{Perturbative Corrections to the Wave Function}

We will now make use of this expression for the Yukawa coupling of the
fermions to the Goldstone mode to estimate the effect that
renormalization has on the wave functions of the fermions.  We will
look at the regime near the IR brane where the Yukawa is growing large
but still perturbative.  We will first consider how the fermion
operators on each slice, and the operators connecting slices, are
renormalized and then use the form of the renormalized parameters to
understand the new fermion wavefunction in the fifth direction.

We first rescale the fermion fields to have canonical kinetic terms.
The Lagrangian we are now working with is:
\begin{eqnarray}
&& -i\left(\psi\sigma^\mu\partial_\mu\bar\psi +
\bar\chi\bar\sigma^\mu\partial_\mu\chi\right)
+ \psi\partial_5\chi - \bar\chi\partial_5\bar\psi + \frac{c}{z}\left(\chi\psi + \bar\psi\bar\chi\right)
\nonumber \\ && \qquad
+\frac{1}{2} \partial_\mu\phi\partial^\mu\phi +
iy(z)\phi\left(\bar\psi\bar\chi - \chi\psi\right)\,.
\label{BareYukawaLagrangian}
\end{eqnarray}
The full action should have boundary terms which come about
from an integration by parts, but these will not be important for the
bulk analysis which we perform here.

\FIGURE[t]{\epsfig{width=.8\textwidth,file=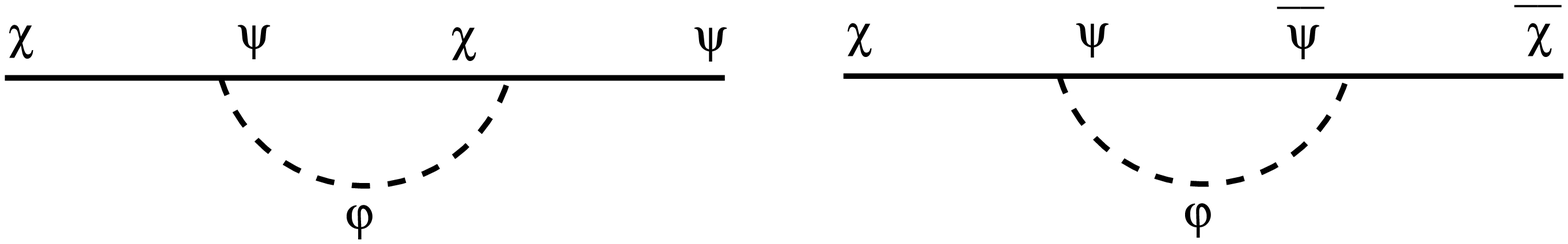}
\caption{The Goldstone exchange contributions.  The left graph gives
rise to mass and $\partial_5$ renomalization while the graph on the
right renormalizes the kinetic term.}
\label{FermiGoldstoneFeynman}}
From evaluation of the diagrams in Fig.~\ref{FermiGoldstoneFeynman}
we find that in the perturbative regime, the form of the corrected
action due to the Yukawa coupling is expected to be
\begin{eqnarray}
&& -i(1+A y(z)^2)\left(\psi\sigma^\mu\partial_\mu\bar\psi +
\bar\chi\bar\sigma^\mu\partial_\mu\chi\right)
+ (1+B y(z)^2) \left( \psi\partial_5\chi - \bar\chi\partial_5\bar\psi\right) 
\nonumber \\
&& + \left(1-By(z)^2\right) \frac{c}{z}\left(\chi\psi + \bar\psi\bar\chi\right) +
B y(z)^2 \partial_z \ln y(z) (\chi\psi - \bar{\psi}\bar{\chi})\,.
\label{correctedYukawaLagrangian}
\end{eqnarray}
The constants $A,B$ would, in perturbation theory, be ${\cal O}
(1/16\pi^2)$.  Since perturbation theory is not applicable here we will
leave them as undetermined constants of order one.\footnote{$A,B$ 
may bring in a dependence on $\ln(R'/R)$ unless the local cutoff,
$v_n$, is used appropriately.}

As stated above, we wish to give a mass of order $1/R'$ to the lightest 
mode of the $\chi$ field (which is localized at the IR brane if 
$c<-\frac{1}{2}$) by coupling that field to a neutral mode on 
the IR brane.  We can, in principle, give this fermion a {\em site
mass} of order $1/R'$, but wavefunction corrections will suppress or 
enhance this mass if the mode is pushed away or towards the IR brane.  In 
addition by rescaling the field to get a canonical kinetic term, the 
effective coupling on the IR brane will be suppressed or enhanced if 
the four dimensional kinetic term is enhanced or suppressed, 
respectively.

Taking these considerations into account and making the
approximation that the Yukawa coupling is linear in $z$ 
(dropping the $\log$ dependence), we find the new zero
mode wave function:
\begin{equation}
g^{(0)}(z) = N_0 z^{-c} \left(1+B \frac{16 \pi^2 }{ b} \frac{z^2 }{ R'^2} 
\right)^{c-1/2}\,,
\end{equation}
where $N_0$ is a normalization constant.  After including the modified
kinetic term in the normalization condition, we find that the
effective mass term on the IR brane which couples the gauge singlet
fermion to our light IR localized mode is
\begin{equation}
\frac{\left(1+\tilde{B}\right)^{c-1/2}}
{R' \sqrt{\int \frac{d\tilde{z}}{\tilde{z}^{2c}}\left(1+\tilde{A}\tilde{z}^2 \right)
\left(1+\tilde{B}\tilde{z}^2 \right)^{2c-1}}}\,,
\label{eq:IRfermionmass}
\end{equation}
where $\tilde z = z/R'$, $\tilde{A}= A\frac{16 \pi^2}{b}$ and similarly for $\tilde{B}$.
We plot this mass in units of $1/R'$ for various values of $\tilde{A}$
and $\tilde{B}$ in Fig.~\ref{FermionMass}.
\FIGURE[t]{\epsfig{width=.6\textwidth,file=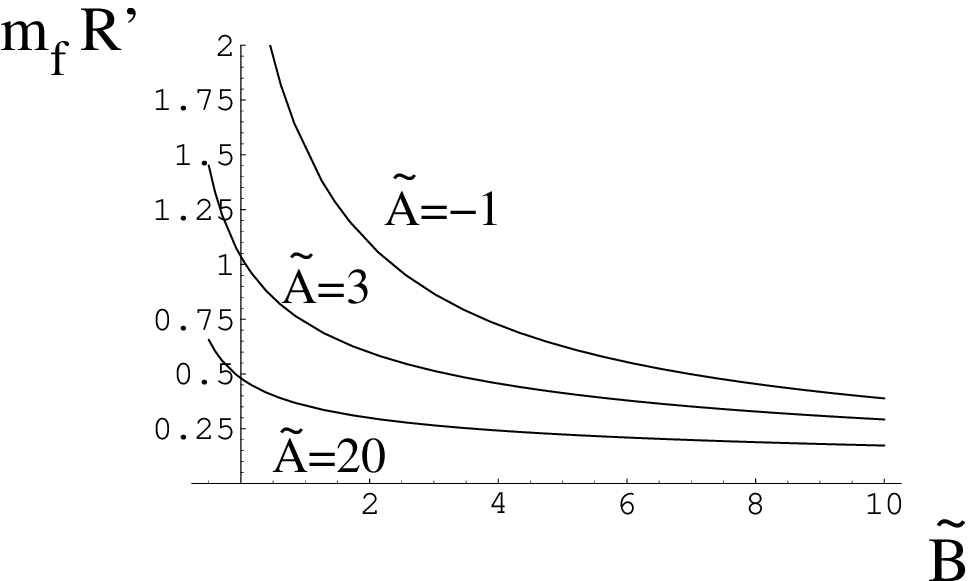}
\caption{The mass of the IR localized fermion in units of the KK scale,
$1/R'$ as a function of $\tilde B$ for three values of $\tilde A$: $-1$,
$3$, $20$ from top to bottom.  We have chosen $c=-1$ and performed
the integral in~(\ref{eq:IRfermionmass}) from $\tilde{z}=0$ to $1$.}
\label{FermionMass}}
For negative values of $\tilde A$ or $\tilde B$ it is less clear that
the perturbative expressions given here can be used to estimate the
mass at the IR brane since there is a pole in either the kinetic term
or the wave function.  Nevertheless, if we take this estimate of the
mass seriously even for large values of $\tilde A$ and $\tilde B$,
then the mass of the unwanted fermion is some fixed fraction of
the KK mass, $1/R'$.  In the large $N$ limit, where the KK modes
decouple, this fermion will still be removed from the low energy
spectrum.  Problems may arise, however, if either $\tilde A$ or
$\tilde B$ grow with $N \sim \ln(R'/R)$ so that the fermion mass is
not a fixed fraction of the KK mass.  This may happen if the lattice
regularization does not respect the local cutoffs, $v_n$, which is a 
non-trivial issue, as we explain in the next section.

A complete 
determination of the actual mass of this fermion mode will have to be done 
using a full non-perturbative lattice computation.

\section{Comments on the Full Regularization of the Theory
  \label{sec:lattice}}
\label{sec:8}
\setcounter{equation}{0}

As discussed in section~\ref{ContinuumWarpedFermions}, the warping is a
crucial feature of the extra dimension.  It is the warp factor which
suppresses the mass of the gauge boson below the scale of the KK modes,
\begin{equation}
M_W^2 \sim \frac{m_{KK}^2 }{ \log \frac{R' }{ R}}\,,
\end{equation}
which is necessary in order to find a limit which restores the gauge
symmetry without making the fermions light.  However, this logarithmic
suppression of the mass appears to be an inefficient way to approach
the symmetric phase and we might hope to do better.  In general, a more
highly warped background will lead to a larger mass suppression and so
we might hope to restore the gauge symmetry more efficiently with a
stronger warp factor than that from AdS.  Only in AdS, however, do we
know of a scaling symmetry which protects the warp factor under
renormalization and without a symmetry it is not clear that a stronger
warping could be maintained.

Unfortunately a na\"\i{}ve lattice regularization does not respect the
scaling symmetry of AdS and will therefore most likely change the wave
functions and masses of the modes we are interested in.  In
particular, if we want to maintain the hypercubic subgroup of the four
dimensional Lorentz group, we only find lattices that have an equal
spacing, $a_4$, in the four flat directions {\em independent} of the
location in the fifth direction, $z$.  For a scalar field the action
is:
\begin{equation}
\frac{1}{2} \sum_{i,j} a_4^4 \delta z_j \left(\frac{R}{z_j}\right)^5
\left[ \left(\frac{z_j}{R}\right)^2 \left( \frac{(\Delta_4 \phi_{ij})^2}{a_4^2}
- \frac{(\Delta_z \phi_{ij})^2}{\delta z_j^2} \right) - m^2 \phi_{ij}^2 \right]\,,
\end{equation}
where $i$ represents a general four dimensional index and $j$ represents
an index along the fifth dimension.  By defining a dimensionless field
\begin{equation}
\tilde{\phi}_{i,j} = a_4 \sqrt{aR}\, w^{j-1} \phi_{i,j},
\end{equation}
we may rewrite the action as
\begin{equation}
\frac{1}{2} \sum_{i,j} \left[ \left(\Delta_4 \tilde\phi_{i,j}\right)^2
- \left(\frac{a_4}{aR}\right)^2w^{2j} \left(\tilde\phi_{i,j+1} - (1+a)\tilde\phi_{i,j}\right)^2
-a_4^2 w^{2j-2} m^2 \tilde\phi_{i,j}^2 \right]\,.
\end{equation}
We can now see that on the lattice, the mass will still be warping
down along the fifth direction.  However, it is the
four dimensional lattice spacing, $a_4$, which sets the scale of the
cutoff and that scale is constant.  Quantum corrections may therefore
destroy the warping of the bulk masses which were necessary to
generate a separation of scales between the low energy fermion and
gauge boson masses.

Of course, an exact AdS lattice should not be necessary, and it is
likely that an appropriate choice of bare parameters would generate an
effective action with the properties needed.  This mild tuning of the
lattice may be the most practical route for constructing a chiral
gauge theory.  However, if the theory proves difficult to tune, it
would be nice to know whether there exists, at least in principle, a
regularization which does respect the scaling symmetry of AdS.

We might find a hint of this using higher order derivative operators
to regulate the theory.  The coefficients of these derivative
operators are dimensionful parameters which \emph{can} be made to
respect the scaling symmetry while maintaining the 4D Lorentz
invariance.  We consider here only a scalar field theory, and the more
involved calculation in a gauge theory will need further
study.

To understand how the higher derivatives respect the scaling symmetry,
we begin by considering operators which are quadratic in the field and
have some arbitrary number of derivatives.  Schematically, this looks
like $\partial_M^n \phi^2$.  The indices must be contracted with the
metric which brings in one factor of $z/R$ for each derivative.
Finally, we must rescale the field to get a canonical four dimensional
kinetic term when we are done, $\phi \to \frac{z}{R}\phi$.  So we
start with
\begin{equation}
\int d^4x dz \frac{R}{z} \left( \left(\partial_M\phi\right)^2 + \dots +
\frac{\partial_M^n \phi^2}{\left(\frac{R}{z}\Lambda\right)^{n-2}}
+ \frac{1}{R\Lambda} \frac{\partial_M^{n-1} \phi^2}{\left(\frac{R}{z}\Lambda\right)^{n-3}}
+ \dots \right)\label{eq:hderiv}
\end{equation}
in our action.  The mass scale $\Lambda$ was added to give this higher
derivative operator the right mass dimension.  The terms with fewer
than $n$ derivatives come from $\partial_z$ acting on the factor of
$z/R$ when the fields are rescaled.  Except for the leading factor,
the dimensionful scales are warping down the way we want, but we still
need to make the $z$ direction discrete.  This will bring in one
factor of $z/R$ from $dz$, removing the unwanted leading factor and
making the four dimensional kinetic operator canonical.  Also, the
derivatives in the $z$ direction lead to a coupling between adjacent
sites with a dimensionful coupling parameter given by
\begin{equation}
\partial_z \phi \to \frac{\phi_{j+1} - \phi_j}{a z_j}\,.
\end{equation}
If we choose the five dimensional cutoff $1/a z_j$ to be
$R\Lambda/z_j$ then, indeed, every dimensionful parameter in our
deconstructed AdS$_5$ theory will be warping the way we want.  It is
known that for scalar interactions, as well as for the nonchiral
fermions that we start with, such a regularization renders the theory
finite to all orders in perturbation theory, and it can be
renormalized in the usual way.

To simulate the theory numerically in the non-perturbative regime, one
can put the slices on a four dimensional lattice with uniform
$z-$independent spacing $a_4$.  This will break the scaling symmetry
at the lattice scale $1/a_4$. A theorem by Reisz~\cite{Reisz},
however, states that for diagrams with a negative lattice divergence,
the continuum limit of the lattice perturbation theory is the same as
the continuum perturbation theory.  It is easy to see that a na\"\i{}ve
discretization of (\ref{eq:hderiv}) belongs to this class.  Applying
the theorem to the matching between the continuum deconstructed and
the lattice regularized theory, one would, therefore, expect that the
renormalized theory, at a fixed $\Lambda$, restores the scale
invariance as $a_4\to0$; and taking $\Lambda\to\infty$ holding
$a_4\Lambda = 0$ then recovers the original scale-invariant lattice
theory.

\section{Consistency with Anomalies and Instantons}
\label{sec:9}
\setcounter{equation}{0}

So far we have not discussed gauge anomalies, however if we tried to
perform our procedure in such a way so as to leave an anomalous light
fermion content, then loop corrections would produce a mass for the
gauge boson that is not removed in the small curvature radius limit.
This effect is well known \cite{Preskill:1990fr}: two back to back
triangle anomalies produce a gauge boson mass at order $g^3$.  It is
this effect that shows that an anomalous gauge theory with an unbroken
gauge symmetry is not a consistent possibility, and thus it is to be
expected that this same effect is what prevents the procedure
described in this paper from constructing an anomalous unbroken gauge
theory.

A consistent latticization of a chiral gauge theory should also make
clear how the instantons of the low energy chiral gauge theory are
produced.  In fact, the chiral fermion measure is complex, and its
phase depends on the gauge field configuration in a non-trivial way
leading to the non-conservation of individual fermion currents in an
instanton background.  This is a non-trivial question since the
't~Hooft operator generated by the instanton of a single chiral theory
contains only left handed fields, while the individual instantons in
every gauge group would generate a 't~Hooft operator with equal number
of left and right handed fermions.  In the absence of a Majorana mass
for the gauge singlet fermions one can, in fact, reproduce neither
the phase of fermion measure, nor such non-perturbative effects of the
chiral gauge theory, since the Dirac measures we start with are real
and there is no way to flip the chiralities. This is in accordance
with the observation that the spectrum from (\ref{eq:fullfermion})
will be chiral only for $m_M\neq 0$. For $m_M\neq 0$, however, one can
chain together several instantons such that the right handed fermions
sticking out from the instanton are all transforming under the last
gauge group, where the gauge symmetry is broken. At that site the
right handed fermions mix with the gauge singlet fermions via the
Higgs VEV. These gauge singlet fermions in turn have Majorana masses
which can flip the chiralities, thus all legs of right handed fermions
can be closed up this way.

As a concrete example, consider constructing a chiral $SU(5)$ gauge
theory with a left-handed ${\bf \bar{5}}$ and ${\bf 10}$ fields. With
a sufficient number of singlet fermions and Higges we can make heavy
all the components of the right-handed ${\bf \bar{5}}$ and ${\bf 10}$
at the final lattice site. Then we see that although we started with a
vector-like theory where there are seperately conserved currents
for ${\bf \bar{5}}$'s and ${\bf 10}$'s, the chain of instantons and
mixing with Majorana singlets reduces the global symmetry, and a
current of ${\bf \bar{5}}$'s can be turned into a current of ${\bf
10}$'s. This is illustrated in Fig.~\ref{fig:AdSInstanton} where we
show how a bunch of $SU(5)$ instantons can be chained together to
generate the 't~Hooft operator of a single chiral $SU(5)$ theory.

\FIGURE[t]{\epsfig{width=\textwidth,file=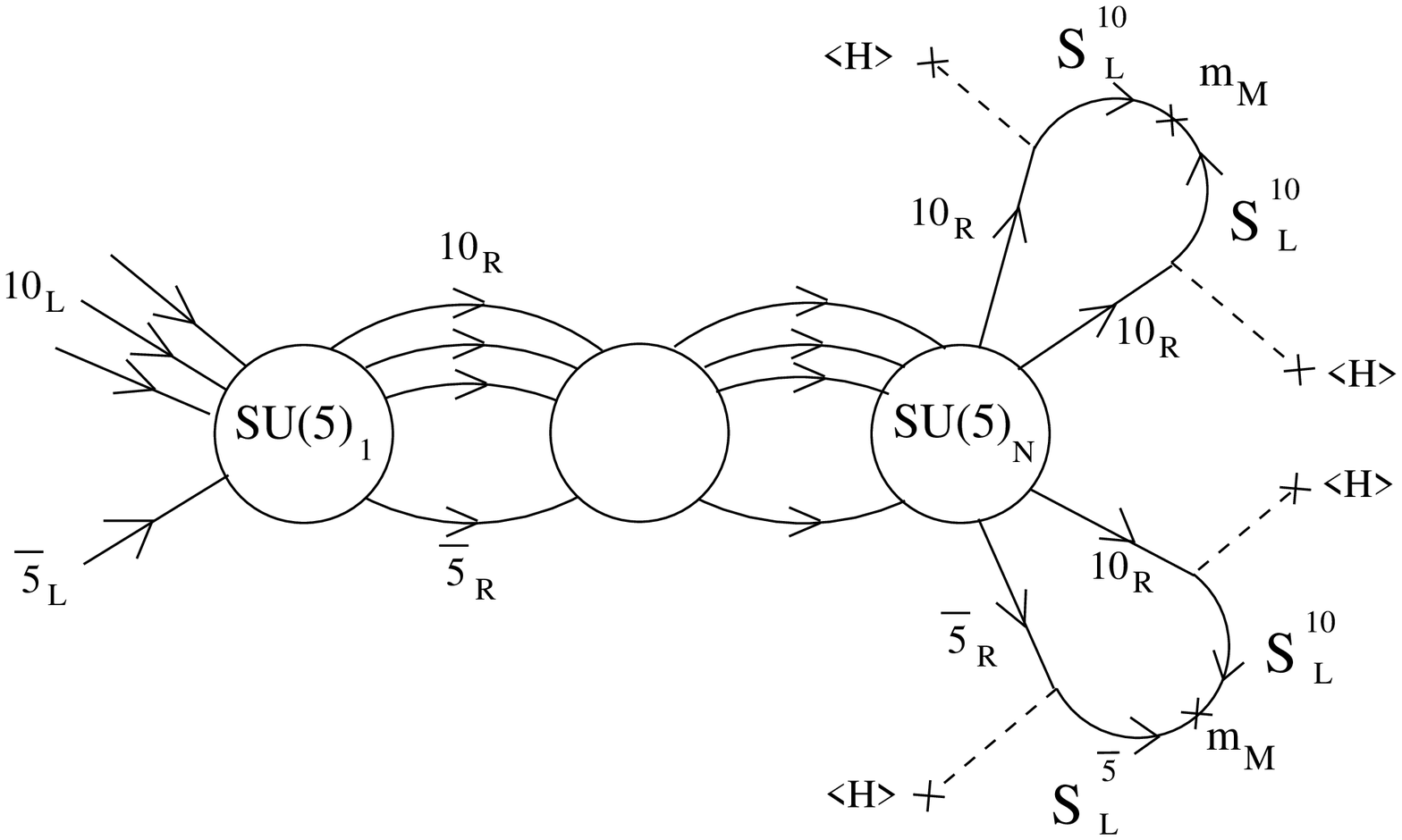}
\caption{A chain of instantons in the extra dimension in an $SU(5)$ theory
with ${\bf \bar{5},10}$. Arrows represent the chirality of the fermion 
field, with left handed fermions going into the instanton, and right 
handed coming out. A Dirac mass connects a left and a right handed fermion
so the flow of the arrow will be continuous, while a Majorana mass reverses
the direction of the arrow, since it connects two left or two right handed 
fields. Note, that there are three zero modes for a ${\bf 10}$ field,
so the 't~Hooft operator is ${\bf 10}^3 {\bf \bar{5}}$ (see~\cite{instantons}
for details).
\label{fig:AdSInstanton}}}

\section{Conclusions}
\label{sec:conclusions}
\setcounter{equation}{0}
We have considered the domain wall fermion construction of chiral
gauge theories in the presence of non-vanishing curvature in the extra
dimension. In the discretized theory without scalar VEV's there are
two light fermions localized at opposite ends of the extra dimension,
and the theory is non-chiral.  The main new feature of these models is
that one can restore gauge invariance in the presence of scalar VEV's
on the IR brane by taking the limit of small curvature radius.  Then
this scalar VEV can be used to remove one of the two fermion
chiralities from the theory. We have checked numerically that the
classical theory will indeed result in a chiral gauge theory. The
chiral instanton operator can also be reproduced in this model. The
main worry is that in the limit of small curvature radius, one light
scalar becomes strongly coupled near the IR brane. This could
potentially result in additional light fermions. In order to find out
whether or not the theory is indeed chiral at the non-perturbative
level, a full lattice simulation needs to be performed.

\section*{Acknowledgments}
\setcounter{equation}{0}

We thank Thomas DeGrand, Marteen Golterman, Erich Poppitz, and Martin Schmaltz for useful
discussions.  T.B., M.M. and Y.S. are supported by the U.S. Department
of Energy under contract W-7405-ENG-36.  C.C. is supported in part by
the DOE OJI grant DE-FG02-01ER41206 and in part by the NSF grants
PHY-0139738 and PHY-0098631.  J.T. is supported in part by DOE grant 
DE-FG03-91ER40674.


\end{document}